\documentclass[showpacs,preprint,aps]{revtex4}
\usepackage{graphicx}
\usepackage{dcolumn}
\usepackage{bm}
\begin{document}
\setcounter{page}{1}
\title
{Eta meson rescattering effects in the 
${\rm p}\,+ \,^6{\rm Li} \, \rightarrow\,  \eta \,+\,^7{\rm Be}$ 
reaction near threshold}
\author
{N. J. Upadhyay$^1$,  
N. G. Kelkar$^2$ and 
B. K. Jain$^1$}
\affiliation
{$^1$ Dept. of Physics, University of Mumbai, Kalina, Mumbai, India\\
$^2$ Dept. de Fisica, Univ. de los Andes,
Cra. 1E, No. 18A-10, Bogota, Colombia
}
\begin{abstract}
The ${\rm p}\, + \,^6{\rm Li} \, \rightarrow\,  \eta \,+\,^7{\rm Be}$ 
reaction has been investigated with an emphasis on the $\eta$ meson and
$^7$Be interaction in the final state. Considering the $^6$Li and $^7$Be 
nuclei to be $\alpha$-d and $\alpha$-$^3$He clusters respectively, the 
reaction is modelled to proceed via the ${\rm p}\, + \, {\rm d} 
[\alpha] \, \rightarrow \, ^3{\rm He} [\alpha] \, +\, \eta$ reaction 
with the $\alpha$ remaining a spectator. The $\eta$ meson interacts with 
$^7$Be via multiple scatterings on the $^3$He and $\alpha$ clusters inside
$^7$Be. The individual $\eta$-$^3$He and $\eta$-$\alpha$ scatterings 
are evaluated using few body equations for the $\eta$-3N and $\eta$-4N 
systems with a coupled channel $\eta$-N interaction as an input.
Calculations including four low-lying states of $^7$Be lead to a 
double hump structure in the total cross section corresponding to 
the $L = 1 \,(J = (1/2)^-, (3/2)^-)$ and $L = 3\, (J = (5/2)^-, (7/2)^-)$ 
angular momentum states. The humps arise due to the 
off-shell rescattering of the $\eta$ meson on the $^7$Be nucleus 
in the final state. 
\end{abstract}
\pacs{25.40.Ve, 21.85.+d, 25.10.+s}
\maketitle

\section{Eta meson interactions with light nuclei}
The past few years have seen extensive investigations of the $\eta$ meson 
producing reactions at close to threshold energies. The experiments 
are aimed either at directly searching \cite{taps} for the possible 
existence of eta-mesic nuclei as a result of the strong attractive 
nature of the $\eta$-N interaction \cite{bhaleliu} or studying the 
final state eta-nuclear interaction to eventually conclude on the 
existence of eta-mesic nuclear states \cite{weme,others}.
A common feature of the data on $\eta$ production near threshold is the 
strong deviation of the cross sections from phase space. It can be 
understood as a manifestation of the strong attractive $\eta$-N interaction 
(arising basically due to the proximity of the $\eta$-N threshold to 
the $S_{11}$ resonance N$^*$(1535)). Experiments on $\eta$ production have 
been performed in nucleon-nucleon collisions and have been extended 
to proton-deuteron and deuteron-deuteron collisions too \cite{exp3He4He}. 
Though the  
focus of reactions such as the p + d $\rightarrow$ p + d + $\eta$ and 
${\rm p}\, + \, {\rm d} \, \rightarrow \, ^3{\rm He}\, +\, \eta$ is on 
investigating possibilities of bound states of eta mesons and 2 - 3 nucleon 
systems, theoretical studies of the reaction mechanism revealed 
interesting features too \cite{we3and4}. In the p-d collisions, the 
production near threshold is found to be dominated by a two-step mechanism 
where the large momentum transfer in producing the $\eta$ meson is shared 
by three nucleons.   
These findings naturally led to the curiosity of what happens when the 
$\eta$ interacts with more than three or four nucleons in the final nucleus. 
With this motivation, measurements of the 
${\rm p}\, + \,^6{\rm Li} \, \rightarrow\,  \eta \,+\,^7{\rm Be}$ reaction 
were carried out by the Turin group in 1993 \cite{turin} at an incident 
energy of 683 MeV. A theoretical study of this reaction along with others
of the type, ${\rm a}\, + \,^6{\rm Li} \, \rightarrow\,  b\,+\,^7{\rm Be}$ 
and ${\rm a}\, + \,^6{\rm Li} \, \rightarrow\,  b\,+\,^7{\rm Li}$ 
was performed in \cite{wilkin}. Part of the emphasis of this work was on
obtaining the right form factors for $^7$Li (and $^7$Be) and the 
interactions of the mesons with the nuclei in the final states were 
neglected. The interest in the 
${\rm p}\, + \,^6{\rm Li} \, \rightarrow\,  \eta \,+\,^7{\rm Be}$ reaction 
revived once again by the recent proposal of studying this reaction at
COSY, J\"ulich, at an incident energy of 673 MeV \cite{cosy}.  

Having performed detailed theoretical studies of the 
p + d $\rightarrow$ p + d + $\eta$ and
${\rm p}\, + \, {\rm d} \, \rightarrow \, ^3{\rm He}\, +\, \eta$ reactions 
\cite{we3and4} and the $\eta$ meson interactions with the deuteron, 
$^3$He and $^4$He nuclei \cite{weme}, 
we now develop a model to study the interaction 
of $\eta$ mesons with a $^3$He-$^4$He cluster, namely, the $^7$Be nucleus. 
The $\eta$-$^7$Be interaction is then incorporated in a theoretical 
calculation of cross sections for the 
${\rm p}\, + \,^6{\rm Li} \, \rightarrow\,  \eta \,+\,^7{\rm Be}$ reaction, 
with four possible low-lying states of the $^7$Be nucleus. 
An analysis of its effects on the 
${\rm p}\, + \,^6{\rm Li} \, \rightarrow\,  \eta \,+\,^7{\rm Be}$ cross 
sections presented here, should be useful in motivating further 
experimental studies of this reaction.  

\section{Cluster model approach}
Based on literature which supports considering the $^7$Be nucleus as a 
bound state of an alpha ($^4$He) and $^3$He \cite{berclust}, we model the 
$\eta$-$^7$Be final state interaction in the form of a 
three body problem of the 
$\eta$-$^3$He-$^4$He interaction. Regarding the $^6$Li too as a cluster of 
an alpha and a deuteron \cite{lithclust}, the 
${\rm p}\, + \,^6{\rm Li} \, \rightarrow\,  \eta \,+\,^7{\rm Be}$ reaction 
is considered to proceed through the 
${\rm p}\, + \, {\rm d} \, \rightarrow \, ^3{\rm He}\, +\, \eta$ reaction 
with the $\alpha$ remaining a spectator. 
Besides, the present work focuses on the low energy region of 
$\eta$ production where (a) the 
${\rm p}\, + \, {\rm d} \, \rightarrow \, ^3{\rm He}\, +\, \eta$ 
production amplitude is large 
(i.e. the p and d interact strongly to produce an $\eta$) 
and (b) the cluster picture of low lying levels of $^7$Be and 
$^6$Li is reasonably good. There exists in principle, the possibility 
of considering the deuteron as a spectator. However, a reaction 
of the type ${\rm p}\, + \, {\alpha} \, \rightarrow \, ^5{\rm Li}\, +\, \eta$ 
followed by the combination of $^5{\rm Li}$ and a deuteron to form the 
$^7$Be nucleus, does not agree with the cluster model approach here 
(since $^7$Be is hardly known to be a cluster of d + $^5$Li, where 
$^5$Li is in fact unstable). The possibility of an intermediate 
${\rm p}\, + \, {\alpha} \, \rightarrow \, {\rm p}\,
+\,\alpha +\, \eta$ reaction,
with further steps of the final state ${\rm p}$ from this reaction combining 
with the spectator deuteron, i.e., 
${\rm p} \,+\,{\rm d}$ forming a $^3$He which eventually
combines with the final state $\alpha$ (from the above reaction) 
to form $^7$Be still remains. However, this is 
not a practical option with no information available on the 
${\rm p}\, + \, {\alpha} \, \rightarrow \, {\rm p}
\,+\,\alpha +\, \eta$ reaction. 

\subsection{Elastic scattering of eta mesons on $^7$Be} 
As mentioned above, we consider the $^7$Be nucleus as a two body system made
up of a $^3$He and $^4$He nucleus and construct an 
elastic transition matrix for 
the three body problem of an 
$\eta$ meson, $^3$He nucleus and an $\alpha$ ($^4$He). 
The individual scattering of the $\eta$ meson on $^3$He and 
$\alpha$ is evaluated using the $t$-matrices constructed earlier 
by the present authors \cite{weme}. These t-matrices are numerically 
evaluated using 
few body equations and include the off-shell rescattering of the $\eta$ 
on the nucleons inside $^3$He and $^4$He. The $\eta$-$^3$He t-matrix 
is tested to reproduce the 
${\rm p}\, + \, {\rm d} \, \rightarrow \, ^3{\rm He}\, +\, \eta$ cross section 
reasonably well \cite{we3and4}. 

To formulate the three body problem of the $\eta$-$^3$He-$^4$He interaction, 
let us define $\vec{r}_1$ and $\vec{r}_2$ to be the coordinates of the 
$^3$He and $^4$He nuclei respectively, 
with respect to the mass-7 centre of mass system. Defining 
the internal Jacobi coordinate of the relative distance between 
the $^3$He and $^4$He nuclei as $\vec{x}$, one can see that 
$\vec{r}_1 = a_1 \vec{x}$ and $\vec{r}_2 = a_2 \vec{x}$ with 
$a_1 = 4/7$ and $a_2 = - 3/7$. The $\eta$-$^7$Be t-matrix is then 
written as, 
\begin{equation}\label{etaBetmat}
T_{\eta-7Be} (\vec{k}^{\prime},\vec{k},z) \,=\,\int\, d^3x\, 
|\Psi^7_L(\vec{x})|^2\, 
\, [ \,T_1(\vec{k}^{\prime}, \vec{k}, a_1 \vec{x}, z)\,+ 
\,T_2(\vec{k}^{\prime}, \vec{k}, a_2 \vec{x}, z)\,] 
\end{equation}
where $T_1(\vec{k}^{\prime}, \vec{k}, a_1 \vec{x}, z)$ and $T_2(\vec{k}^{\prime}, 
\vec{k}, a_2 \vec{x}, z)$ are 
the medium modified t-matrices for the off-shell $\eta$ scattering on 
the bound $^3$He and $^4$He respectively. 
$\Psi^7_L(x)$ represents the cluster wave 
function of $^7$Be with angular momentum $L$ 
and $z \,=\, E \,-\,|\varepsilon_0|\,+\,i\epsilon$, where, $E$ 
is the total $\eta$-nucleus energy in the centre of mass 
and $\varepsilon_0$ is the 
energy required for the break up of $^7$Be $\to\,^3$He+$^4$He. 
The in-medium
$\eta$-$^3$He and $\eta$-$^4$He t-matrices are written using a Faddeev 
type decomposition \cite{rakit}, namely,
\begin{equation}\label{eta3He4He} 
T_i(\vec{k}^{\prime}, \vec{k}, a_i \vec{x}, z)\, =\, t_i(\vec{k}^{\prime}, 
\vec{k}, a_i \vec{x}, z)\,+ \, 
\int\, {d\vec{k}^{\prime \prime} \over (2 \pi)^3}\, \, 
{t_i(\vec{k}^{\prime}, \vec{k}^{\prime \prime}, a_i \vec{x}, z) \over z \,-\, 
k^{\prime \prime \,2}/2\mu } \,\,\sum_{j \ne i} \,
T_j(\vec{k}^{\prime \prime}, \vec{k}, a_j \vec{x}, z)\, ,
\end{equation} 
where $i = 1, 2$ and the indices $1$ and $2$ correspond to the $^3$He 
and $^4$He t-matrices respectively. The $t_i$'s represent the single scattering
terms and are the matrices for purely elastic $\eta$-$^3$He and 
$\eta$-$^4$He scattering. They have the form, 
\begin{equation}\label{tetan}
t_i(\vec{k^\prime},\, \vec{k}\,;\vec{r_i}\,;z) =
t_i(\vec{k^\prime},\,
\vec{k}\,;z)\, exp [\,i (\, \vec{k} -
\vec{k^\prime}\,)\cdot\,\vec{r_i}\,] \, , 
\end{equation}
with $\vec{r}_i = a_i \vec{x}$ and $i = 1, 2$ as mentioned above.  
At the low energies considered here, the $\eta$-N interaction is
dominated by the $S_{11}$ resonance N*(1535) 
and hence we perform a partial wave 
expansion and retain only $s$-waves, which reduces (\ref{eta3He4He}) to 
the following two equations for $\eta$-$^3$He ($T_1$) and $\eta$-$^4$He 
($T_2$): 
\begin{equation}\label{eta3He4He2} 
T_1({k}^{\prime}, {k}, a_1 {x}, z)\, =\, t_1({k}^{\prime}, {k}, a_1 {x}, z)\,
+ \, \int\, {d{k}^{\prime \prime} \over 2 \pi^2}\,{k}^{\prime \prime \,2} \, 
{t_1({k}^{\prime}, {k}^{\prime \prime}, a_1 {x}, z) \over z \,-\, 
k^{\prime \prime \,2}/2\mu } \,
T_2({k}^{\prime \prime}, {k}, a_2 {x}, z)
\end{equation}
\begin{equation}\label{eta3He4He3}
T_2({k}^{\prime}, {k}, a_2 {x}, z)\, =\, t_2({k}^{\prime}, {k}, a_2 {x}, z)\,
+ \, \int\, {d{k}^{\prime \prime} \over 2 \pi^2}\,{k}^{\prime \prime \,2} \, 
{t_2({k}^{\prime}, {k}^{\prime \prime}, a_2 {x}, z) \over z \,-\, 
k^{\prime \prime \,2}/2\mu } \,
T_1({k}^{\prime \prime}, {k}, a_1 {x}, z)\, , 
\end{equation} 
where $t_i(\vec{k^\prime},\, \vec{k}\,;\vec{r_i}\,;z)$ in 
Eq. (\ref{tetan}) has reduced to  
$ t_i({k}^{\prime}, {k}, a_i {x}, z)$ written in terms of the spherical 
Bessel functions $j_0(a_i\, x\, k)$ and $j_0(a_i\, x\, k^{\prime})$. 
Replacing the equation for $T_2$ in $T_1$, 
we obtain a recursive relation for $T_1$ as follows,
\begin{eqnarray}\label{eta3He} 
T_1({k}^{\prime}, {k}, a_1 {x}, z)\, =\, t_1({k}^{\prime}, {k}, a_1 {x}, z)\,
+ \, \int\, {d{k}^{\prime \prime} \over 2 \pi^2}\,{k}^{\prime \prime \,2} \, 
{t_1({k}^{\prime}, {k}^{\prime \prime}, a_1 {x}, z) \over z \,-\, 
k^{\prime \prime \,2}/2\mu } \,
t_2({k}^{\prime \prime}, {k}, a_2 {x}, z)\\ \nonumber
+\,\int \,\int\, {d{k}^{\prime \prime} d{\tilde k}
\over 4 \pi^4}\,{k}^{\prime \prime \,2} \,{\tilde k}^2\,  
{t_1({k}^{\prime}, {k}^{\prime \prime}, a_1 {x}, z) \,
t_2({k}^{\prime \prime}, {\tilde k}, a_2 {x}, z)\, 
T_1({\tilde k}, {k}, a_1 {x}, z)
\over (z \,-\, k^{\prime \prime \,2}/2\mu) \,(z \,-\,{\tilde k}^2/2\mu)} \,.
\end{eqnarray}
Once $T_1$ is evaluated from (\ref{eta3He}), 
it can be replaced into the equation for $T_2$ 
and the two can be substituted in (\ref{etaBetmat}) to evaluate the 
$\eta$-$^7$Be t-matrix. Thus (\ref{etaBetmat}) is evaluated retaining only 
$s$-waves in $T_1$ and $T_2$.  
$T_1$ is evaluated numerically using the 
$\eta$-$^3$He and $\eta$-$^4$He t-matrices, $t_1$ and $t_2$ respectively, 
as inputs. $t_1$ and $t_2$ are themselves evaluated numerically using 
few body equations and an input coupled channel $\eta$-N t-matrix. 
Details of this formalism can be found in \cite{weme}. The two models 
of the elementary $\eta N$ t-matrix used here will be discussed in the next 
subsection. 

\subsubsection{Models of elementary $\eta$-nucleon scattering} 
The coupled channel $\eta$-N t-matrix which is required for the 
evaluation of the $\eta$-$^3$He and $\eta$-$^4$He t-matrices, $t_1$ and $t_2$, 
is taken from two different models available in literature. 
In \cite{fix}, a transition matrix including the $\pi$N and
$\eta$N channels with the N*(1535) resonance playing a
dominant role was constructed. It consisted of the
meson - N* vertices and the N* propagator as given below:
\begin{equation}\label{tfix}
t_{\eta \, N \, \rightarrow \, \eta \, N} (\, k^\prime, \, k; z) = 
{ { \rm g}_{_{N^*}}\beta^2 \over (k^{\prime\,2} +
\beta^2)}\,\tau_{_{N^*}}(z)\,{ {\rm g}_{_{N^*}}\beta^2 \over (k^2 + \beta^2)}
\end{equation}
with,
$\tau_{_{N^*}}(z) = ( \, z - M_0- \Sigma_\pi(z) - 
\Sigma_\eta(z) + i\epsilon \, )^{-1}$,
where $\Sigma_\alpha(z)$ $(\alpha = \pi, \eta)$ are the self energy
contributions from the $\pi N$ and  $\eta N$ loops.
In \cite{fix} elastic and inelastic $\eta$-deuteron scattering as well as
$\eta$ photoproduction on the deuteron was studied using this $\eta N$ model. 
We shall use two parameter sets available, 
one which leads to a scattering length of $a_{\eta N}$ = (0.75, 0.27) fm 
and another which leads to $a_{\eta N}$ = (0.88, 0.41) fm. We shall refer 
to this model henceforth as Model A. 

Model B used in the present work is taken from \cite{wycgre}. The 
t-matrix for $\eta N \to \eta N$ is written in a separable form as, 
\begin{equation}\label{green1}
t_{\eta \, N \, \rightarrow \, \eta \, N} (\, k^\prime, \, k; z)\, = \, 
v(k^\prime) \, t_{\eta \eta}(E) \, v(k) \, ,
\end{equation} 
where the on-shell part, $t_{\eta \eta}(E)$, is 
described in an effective range approximation as, 
\begin{equation}\label{green2}
t_{\eta \eta}(E)^{-1} \, +\, i\,q\,v(q)^2\, =\,{1 \over a} \,+ \, 
{r_0 \over 2} \, q^2\, +\, s\, q^4\, .
\end{equation} 
The off-shell form factors have the form, $v(k) \, =\, 1/( 1\,+\,k^2\, 
\Lambda^2)$, where $\Lambda$ is the length parameter in this model. 
The parameter sets in this model are obtained from a fit to the 
$\pi N \to \pi N$, $\pi N \to \eta N$, 
$\gamma N \to \pi N$ and $\gamma N \to \eta N$ data. 
The parameters required in Eqs. (\ref{green1}) and (\ref{green2}) 
above can be found in \cite{wycgre}. 
We choose four parameter sets with $\eta N$ scattering lengths of 
(0.88,0.25) fm, (0.77, 0.25) fm, (0.51,0.26) fm and (0.4, 0.3) fm. 

\subsubsection{Kinematics of the many body problem} 
The choice of kinematic variables in a 
multiple scattering formalism where the many body 
scattering matrix is written in terms of a two body matrix is not unique in 
literature. The two most commonly used approaches are (i) the fixed 
scatterer approximation (FSA) and (ii) the on-energy shell impulse 
approximation (OEI) \cite{kujaw} (related to yet another approach, namely, 
the fixed impulse approximation (FIA) \cite{crespo}). The difference between 
these approaches (which becomes important at intermediate and high energies) 
lies in the fact that in the FSA, the struck nucleon 
is assumed to recoil with the target as a whole, while in the OEI, it 
recoils freely. This means that in the FSA, the two-body operator does not
follow from the equation for a free two-body t-matrix, but rather contains 
the mass of the nucleus (in which the two-body system is embedded) in 
the kinetic energy. Though this seems to be mathematically sound, such a 
two body t-matrix can have physically undesired 
features. For example, in \cite{bely3} 
it was found from a phase shift study that such a two-body t-matrix 
may not display
the resonant behaviour which it would be expected to. In the same work, 
in connection with $\pi$-nucleus scattering, 
it was found that considering a free two-body matrix simulated the 
contribution of continuum states (otherwise neglected in that work) and
brought theory in closer agreement with data. 

Though the differences arising from the particular approaches used above
may not be crucial at the low energies considered in the present work, the 
many-body problem considered here is a bit more complicated and hence 
a small discussion is in order.  
In the works mentioned above, one studies the differences of the 
approaches involved, in a multiple scattering of a hadron on 
the individual nucleons in the nuclear target. Here however, the problem
appears to be that of one multiple scattering problem embedded inside another. 
The $\eta$ mesons scatter off the $^3$He and $^4$He nuclei as in a 
three body multiple scattering problem (of the $\eta$-$^3$He-$^4$He system). 
However, the individual $\eta$-$^3$He and $\eta$-$^4$He scatterings 
are represented by t-matrices for the multiple scatterings
of the $\eta$ on the three and four nucleons inside $^3$He and $^4$He. 
We choose then to work in a framework where we start by evaluating the 
momenta $k$ and $k^{\prime}$ in the $\eta$-$^7$Be centre of mass frame and 
then evaluate the individual $\eta$-$^3$He and $\eta$-$^4$He t-matrices 
at a Lorentz boosted energy-momentum 
in the $\eta$-$^3$He and $\eta$-$^4$He centre of mass systems. 
The energy in the in-medium propagators (inside $^7$Be) is however taken
to be in the $\eta$-$^7$Be centre of mass system. 

\subsubsection{Cluster wave functions of $^6$Li and $^7$Be} 
Since the energy spacing between the first four low-lying levels of the
$^7Be$ nucleus is small, we include the contribution from these four 
levels. We consider the angular momentum states with $L = 1$ and 
$L = 3$ corresponding to the $J = (\,{3/2}^-\,,\,{1/2}^-)$ and 
$J = (\,{7/2}^-\,,\,{5/2}^-)$ levels respectively. 
The cluster wave functions for $^7$Be and 
those for $^6$Li required in the production amplitude of the 
${\rm p}\, + \,^6{\rm Li} \, \rightarrow\,  \eta \,+\,^7{\rm Be}$ reaction 
are (1) generated using a Wood-Saxon potential \cite{khal} and 
(2) taken from a Green's function Monte Carlo
(GFMC) variational calculation with the Urbana potential \cite{for}.
In the GFMC case, the wave function for $^7$Li is used to represent the
$^7$Be one. 
The $\alpha$ and deuteron cluster in $^6$Li is assumed to be in the 
$L = 0$ state. 
\subsection{Production mechanism of the 
${\rm p}\, + \,^6{\rm Li} \, \rightarrow\,  \eta \,+\,^7{\rm Be}$ reaction} 
Assuming that the beam proton interacts with a loosely bound deuteron 
in the $^6$Li nucleus to produce an $\eta$ meson and $^3$He (with 
the $\alpha$ remaining a spectator) the production amplitude for the 
${\rm p}\, + \,^6{\rm Li} \, \rightarrow\,  \eta \,+\,^7{\rm Be}$ reaction 
can be written in terms of that for the 
${\rm p}\, + \, {\rm d} \, \rightarrow \, ^3{\rm He}\, +\, \eta$ process.
The off shell $\eta$ meson thus produced then 
rescatters on $^7$Be (i.e. the $^3$He-$^4$He cluster). 
This production of the $\eta$ 
and its final state interaction (FSI) with $^7$Be is 
represented schematically in Fig. 1, and the corresponding 
transition matrix is written as,
\begin{eqnarray}\label{prodtmat1}
T&=&\langle\,\vec{k_{\eta}}\,,\,m_7\,|\,T_{p\,^6Li\,\rightarrow\,\eta\,^7Be}
\,|\,\vec{k_p}\,;\,m_p\,,\,m_6\rangle\\
\nonumber
&&+\,\sum_{m_7^\prime}\int {d\vec{q}\over(2\,\pi)^3}\,{\langle\,\vec{k}_{\eta}
\,,\,m_7\,|\,T_{\eta\,^7Be \to \eta\,^7Be}
\,|\,\vec{q}\,,\,m_7^\prime\,\rangle \over
E(k_{\eta})\,-\,E(q)\,+\,i\,\epsilon}\,\langle\,\vec{q}\,,\,m_7^\prime\,|\,
T_{p\,^6Li\,\rightarrow\,\eta\,^7Be}\,|\,\vec{k_p}\,;\,m_p\,,\,m_6\rangle
\end{eqnarray}
where $\vec{k_p}$ and $\vec{k_{\eta}}$ are the initial and final momenta in
the centre of mass system. $m_p$, $m_6$ and $m_7$ are the spin projections of 
the proton, $^6Li$ and $^7Be$ respectively. 
\begin{figure}[h]
\includegraphics[width=7cm,height=5cm]{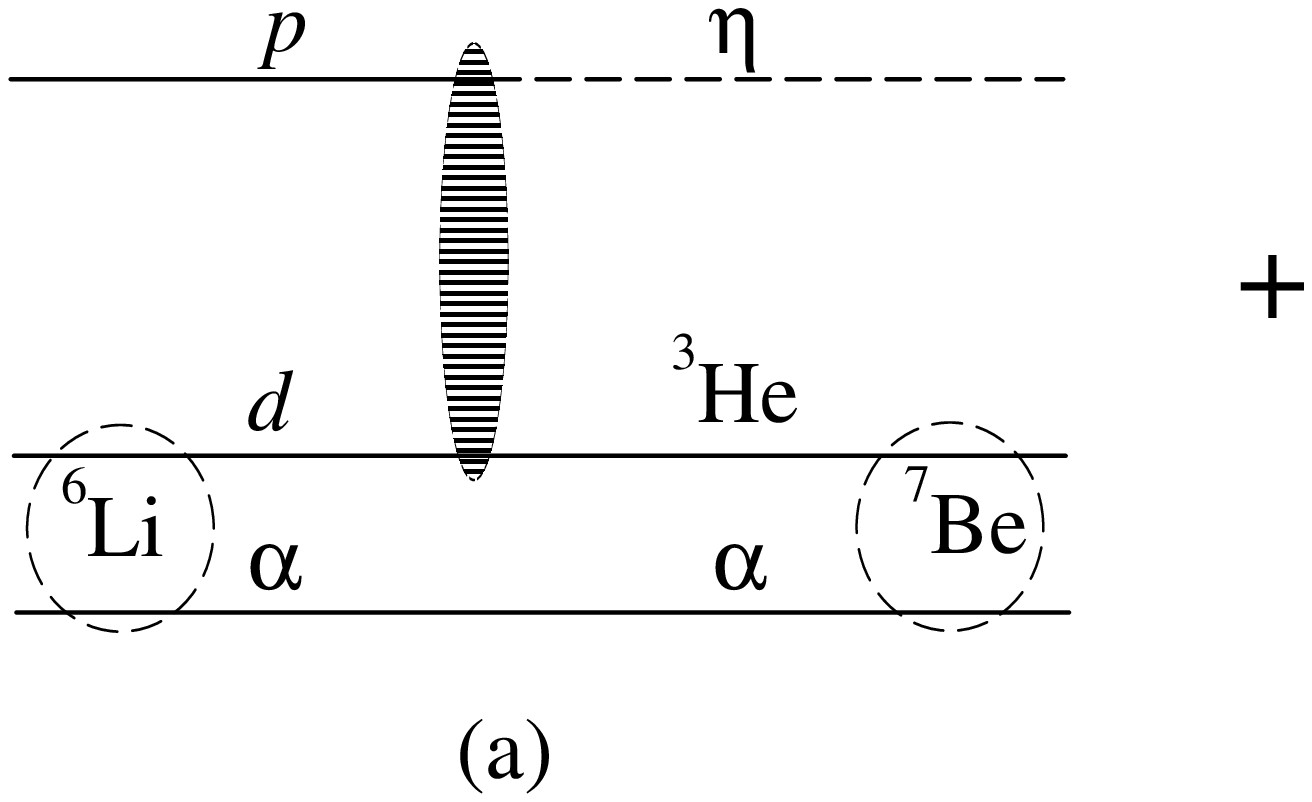}
\includegraphics[width=7cm,height=5cm]{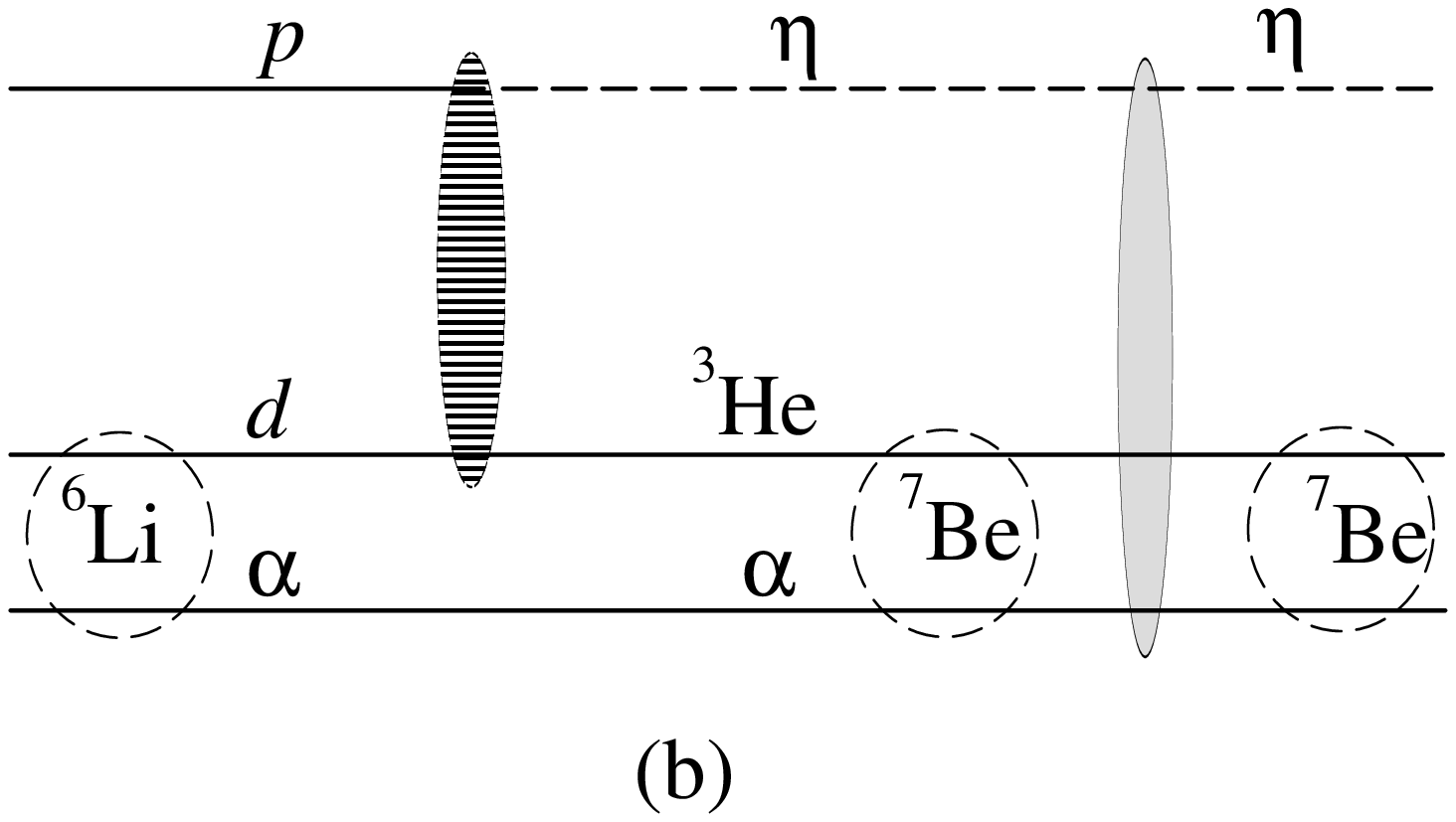}
\\
\caption{\label{fig1:eps1}
Cluster model for $\eta$ production in the 
${\rm p}\, + \,^6{\rm Li} \, \rightarrow\,  \eta \,+\,^7{\rm Be}$ reaction.
Diagram (a) corresponds to the direct on-shell $\eta$ production 
and (b) to an $\eta$ which is first produced off-shell and rescatters via the
$\eta$\, + \,$^7$Be $\,\to \,\eta$ \, +\, $^7$Be process to become on-shell. 
}
\end{figure}
The production matrix for a relative angular momentum
$L$ between the $^3He$ and $\alpha$ is written as, 
\begin{eqnarray}\label{prodtmat2}
\langle\,\vec{k}_{\eta}\,(\vec{q}),\,m_7^\prime\,|\,
T_{p\,^6Li\,\rightarrow\,\eta\,^7Be}\,|\,\vec{k_p}\,;\,m_p\,,\,m_6\rangle
\,= \,{i \over (2 \pi)^3} \,{1 \over \sqrt{4 \pi} } \, \, \int \, 
P_1^2\, dP_1 \, d\Omega_{P_1} \, \Psi^6_0 (P_1) 
\, \,\, \,\,\,\,
\, \,\, \,\,\,\,
\, \,\, \,\,\,\,
\\ \nonumber
\sum_{M, \mu} \, \langle\, J,\, m_7^{\prime}\, |\, {1\over 2},\, \mu,\,L,\, 
M\,\rangle\,\, \,
\Psi^{*\,7}_L ( P_2 )\,\,\,Y^*_{L M}(\hat{P}_2)
\, \,\, \,\,\,\,
\, \,\, \,\,\,\,
\, \,\, \,\,\,\,
\\ \nonumber
\times \, 
\langle\,\vec{k}_{\eta}\,(\vec{q}) 
,\,{-3 \over 7}\vec{k}_{\eta}\,(\vec{q}) + \vec{P}_2,\,
{1 \over 2},\,\mu\,|\,T_{p\,d\,\rightarrow\,\eta\,^3He}\, |\, {1\over 2}, 
\, m_p,\,1, \, m_6,\, \vec{k}_p,\, -{1 \over 3}\vec{k}_p + \vec{P}_1\, 
\rangle \, ,
\end{eqnarray}
where, $\vec{k}_p$ is the momentum of the incoming proton and 
$-[2/6]\vec{k}_p + \vec{P}_1$ of the deuteron in $^6$Li. 
$\mu$ is the spin projection of $^3$He. 
The on-shell $\eta$ meson momentum is denoted as $\vec{k}_{\eta}$ and 
the off-shell one as $\vec{q}$. 
$[-3/7]\vec{k}_{\eta} + \vec{P}_2$ or $[-3/7]\vec{q} + \vec{P}_2$
is hence the momentum of on- or off-shell $^3$He in $^7$Be. 
Since the $\alpha$ particle remains a spectator, 
its momentum in $^6$Li and $^7$Be is required to be the same and 
$-[4/6]\vec{k}_p - \vec{P}_1 \, =\, -[4/7]\vec{k}_{\eta} (\vec{q}) 
- \vec{P}_2$ (for on-shell (off-shell) $\eta$ production). 
Thus $\vec{P}_2\, =\, \vec{P}_1\,+\,[2/3]\vec{k}_p\,-\,[4/7]\vec{k}_{\eta}
 (\vec{q})$ 
(where $\vec{P}_1$ and $\vec{P}_2$ are the Fermi momenta inside 
$^6$Li and $^7$Be respectively). 
The integration in (\ref{prodtmat2}) should in principle have 
been over both the Fermi momenta, $\vec{P}_1$ and 
$\vec{P}_2$, however, the above relation between them renders 
the integration over $\vec{P}_2$ in (\ref{prodtmat2}) redundant. 
Further, when one evaluates the unpolarized cross sections, one sums 
over the spins in the final state and averages over those in the initial
state. As a result, in such a calculation, 
some sums in (\ref{prodtmat1}) and (\ref{prodtmat2}) become redundant. 

The $T-$matrix for the process,
$\langle\,|\,T_{p\,d\,\rightarrow\,\eta\,^3He}\,|\,\rangle$,
is written in a two-step model from our earlier work \cite{we3and4}.
Considering the complexity of the present calculations which include the 
off-shell rescattering as given by the second term in (\ref{prodtmat1})
with the $\eta$-$^7$Be FSI and the fact that the two-step model of 
the ${\rm p}\, + \, {\rm d} \, \rightarrow \, ^3{\rm He}\, +\, \eta$ 
is itself quite involved, we neglect the effect of Fermi
motion on $\langle\,|\,T_{p\,d\,\rightarrow\,\eta\,^3He}\,|\,\rangle$
and hence take it out of the integral over $P_1$ in (\ref{prodtmat2}). 
The momentum space wave functions are expressed in terms of Fourier 
transforms of their radial forms. Thus the integral in momentum space 
is transformed to that in coordinate space.  
All this simplifies Eq. (\ref{prodtmat2}) to a good extent and 
it can be written as,
\begin{eqnarray}\label{equation8}
\nonumber
\langle\,...\,|\,T_{p\,^6Li\,\rightarrow\,\eta\,^7Be}\,|\,...\,\rangle
&=&i^{(L\,+\,1)}\,\sqrt{4\,\pi}\,\sum_{M\,\mu}Y^*_{L\,M}(\hat{Q})\,
F_L(Q)\,
\langle\, J,\, m_7^{\prime}\, |\, {1\over 2},\, \mu,\,L,\,M\,\rangle 
\\
&&\times\,\langle\,...\,|\,T_{p\,d\,\rightarrow\,\eta\,^3He}\,
|\,...\,\rangle
\end{eqnarray}
where, 
\begin{equation}\label{formfac}
F_L(Q)\,=\,\int_0^\infty\,r^2\,dr\,\,\Psi^{*\,7}_L(r)\,j_L(Q\,r)\,
\Psi^6_0(r)\,
\end{equation}
\begin{figure}[h]
\includegraphics[width=7cm,height=7cm]{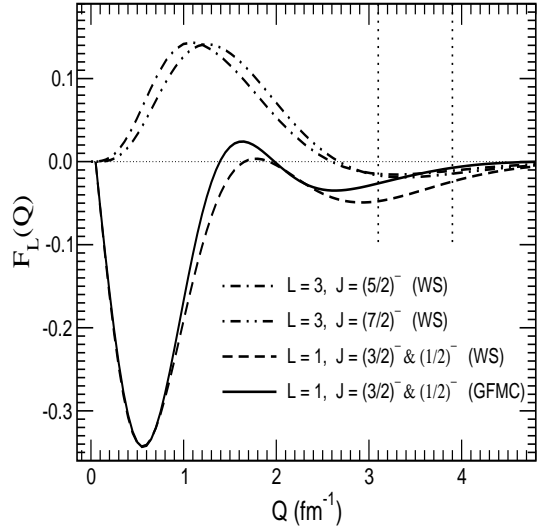}
\caption{\label{fig2:eps1}
The $^6$Li-$^7$Be transition form factor for angular momentum states 
with $L = 1$ and $L = 3$. The solid line corresponds to the 
GFMC variational wave functions with the Urbana potential 
for $L = 1$ and
the dashed, dot-dashed and double dot-dashed lines are  
those generated using a Woods-Saxon potential (for $L = 1$ and $L = 3$).} 
\end{figure}
is the transition form factor for $^6$Li $\to$ $^7$Be, with 
momentum transfer 
$\vec{Q}\,=\,{4\over7}\vec{q}\,-\,{2\over3}\vec{k_p}$ for example in
the off-shell case. Though not written explicitly, the transition 
form factor depends on the total angular momentum $J$ since the 
radial wave function in $^7$Be depends, even if mildly, on $J$.  
In Fig. 2 we present the two form factors with $L=1$ 
($J = 1/2^-, 3/2^-$) and $L=3$ ($J = 5/2^-, 7/2^-$) 
required in the present work, using two different prescriptions 
of the nuclear wave functions mentioned in the previous section. 
With the $L = 1$ levels, $J = 1/2^-$ and $J = 3/2^-$ being very close
to each other, the difference between the two $L = 1$ form factors
is not visible in the figure.  
The form factor enters (\ref{prodtmat1}) via (\ref{equation8}) 
as (\ref{prodtmat2}) is simplified to (\ref{equation8}) due to the neglect of
the Fermi motion in the 
${\rm p}\, + \,{\rm d} \, \rightarrow\, ^3{\rm He} + \eta$ t-matrix. 
It is evaluated at 
$\vec{Q}\,=\,{4\over7}\vec{k}_{\eta}\,-\,{2\over3}\vec{k_p}$ in the first 
term of (\ref{prodtmat1}) whereas over a range of momenta in the 
second term of (\ref{prodtmat1}) where it appears inside the integral. 
The dotted vertical lines in Fig. 2 indicate the relevant range corresponding 
to the beam energies of the present work. We shall see later how this 
difference between the two types of form factors affects the cross sections. 

\section{Results and Discussions} 
In what follows, we shall present the cross section calculations for 
the ${\rm p}\, + \,^6{\rm Li} \, \rightarrow\,  \eta \,+\,^7{\rm Be}$ reaction 
with an emphasis on the $\eta$-$^7$Be FSI. 
The transition matrices for the elementary processes
$\pi$ + N $\rightarrow \, \eta$ + N and $\eta$ + N $\rightarrow \, \eta$ + N 
which enter as inputs to the production and FSI matrices are chosen from 
coupled channels calculations. 
The calculations in Figs 3-7 are done within Model A 
with a choice of parameters which 
corresponds to a scattering length of $a_{\eta N} = 0.88 + i \,0.41$ fm. 
Within the two-step model of the 
${\rm p}\, + \, {\rm d} \, \rightarrow \, ^3{\rm He}\, +\, \eta$ reaction 
we use here, the data on this reaction were reproduced well with this 
choice of the scattering length.

In Fig. 8, a comparison of the total (inclusive) cross sections within models 
A and B of the $\eta N \to \eta N$ t-matrix 
and using different sets of scattering lengths is made. 
\begin{figure}[h]
\includegraphics[width=10cm,height=10cm]{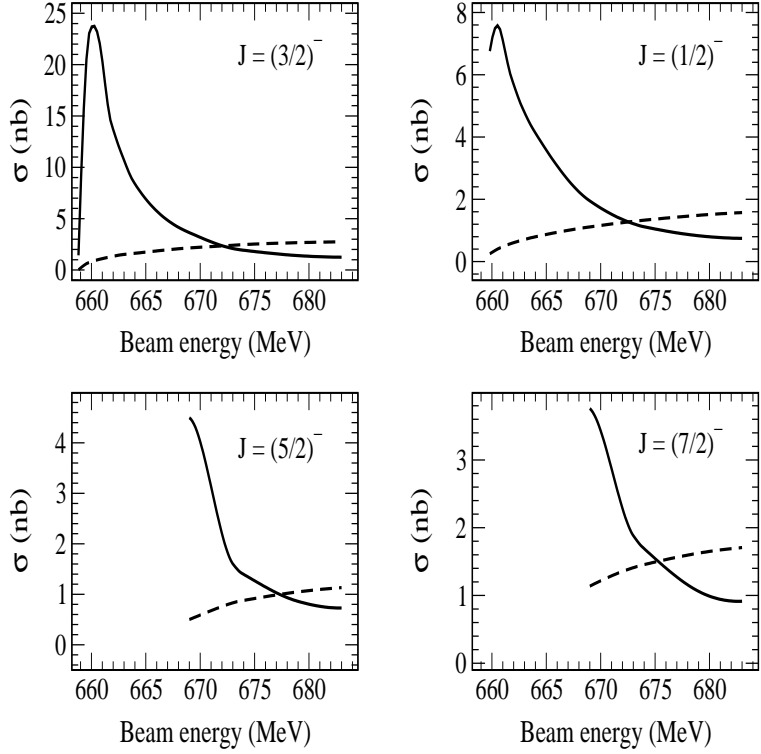}
\caption{\label{fig3:eps1}
Total cross sections as a function of the proton beam energy for different 
states of the $^7$Be nucleus. The dashed lines are plots without the 
inclusion of the $\eta$-$^7$Be final state interaction. The wave functions
for $^6$Li and $^7$Be are generated using the Woods-Saxon potential.}
\end{figure}
\begin{figure}
\includegraphics[width=10cm,height=5cm]{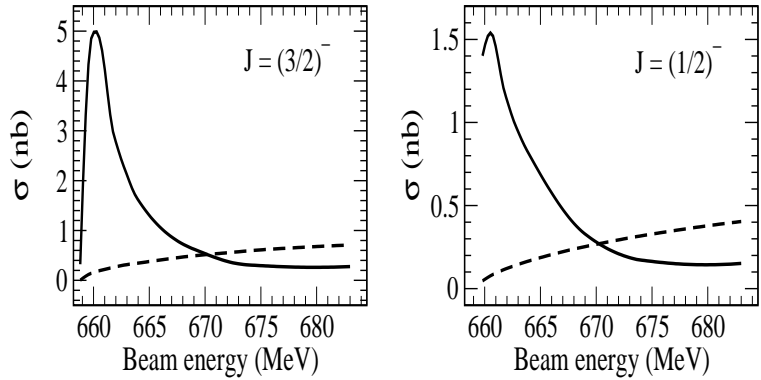}
\caption{\label{fig4:eps1}
Same as Fig. 3 except for the fact that the wave functions
for $^6$Li and $^7$Be are from the GFMC variational method with 
the Urbana potential.}
\end{figure}
\begin{table}[h]
\caption{Beam and excess energies for different levels of the 
$^7$Be nucleus}
\begin{center}
\begin{tabular}{|l|l|l|l|l|}
\hline
Beam energy&\multicolumn{4}{l|}{Excess energy = E$_{\eta-7Be}$\,\,-\,\,
M$_{7Be}$\,\,-\,\,M$_{\eta}$ \,(MeV)}\\
\cline{2-5}
 & \multicolumn{2}{l|}{\rm L = 1} & \multicolumn{2}{l|} 
{\rm L = 3} \\
\cline{2-5}
(MeV) &J = (3/2)$^-$&J = (1/2)$^-$&J = (7/2)$^-$&J = (5/2)$^-$ \\
&ground state&&&\\
\hline
\cline{1-5}
658.8&0.0003&-&-&-\\
\hline
659.8&0.791&0.361&-&-\\
\hline
663.8&3.954&3.524&-&-\\
\hline
669&8.056&7.626&3.426&1.376\\
\hline
673.8&11.855&11.425&7.225&5.175\\
\hline
683&19.11&18.679&14.478&12.428\\
\hline
\end{tabular}
\end{center}
\end{table}
 
\subsection{The 
${\rm p}\, + \,^6{\rm Li} \, \rightarrow\,  \eta \,+\,^7{\rm Be}$ cross 
sections for different $^7$Be levels}
As mentioned earlier, the ${\rm p}\, + \,^6{\rm Li} \, \rightarrow\,  
\eta \,+\,^7{\rm Be}$ reaction is studied with four possible final states 
of the $^7$Be nucleus. The proton beam energies are chosen to study the 
cross sections up to excess energies of about 20 MeV above threshold. 
Due to the differences between the masses of the various $^7$Be levels, 
the threshold for the reaction corresponding to each level differs. 
In Table I, as an example, we list some of 
the beam energies at which we evaluate the cross sections, 
along with the corresponding excess
energies in order to facilitate the understanding of the plots later. 

In Figs 3 and 4, the angle integrated total cross sections for the 
${\rm p}\, + \,^6{\rm Li} \, \rightarrow\,  \eta \,+\,^7{\rm Be}$ reaction 
are shown as a function of the proton beam energy. The dashed lines 
are the cross sections evaluated using only Fig. 1(a) 
corresponding to the first term in Eq. (\ref{prodtmat1}). As is evident 
from these plots, inclusion of Fig. 1(b), i.e., the 
rescattering of the $\eta$ meson with the $^7$Be nucleus (or in other
words the final state $\eta$-$^7$Be interaction) drastically affects 
the shape and magnitude of the cross sections near threshold. 

\subsubsection{The off-shell rescattering contribution} 
The second term in Eq. (\ref{prodtmat1}) which corresponds to rescattering 
(Fig. 1(b)), consists in principle of  
the scattering of on- as well as off-shell $\eta$ mesons on $^7$Be. 
As is generally expected at low energies, 
Fig. 5 clearly shows that neither the plane wave scattering (Fig. 1a) 
nor the pole term in Eq. (\ref{prodtmat1}) (the 
on-shell $\eta$-$^7$Be rescattering) is responsible for the 
near threshold cross section hump. It is the principal value of the integral
in the second term which corresponds to the scattering of off-shell eta 
mesons produced in the 
${\rm p}\, + \,^6{\rm Li} \, \rightarrow\,  \eta \,+\,^7{\rm Be}$ process 
on $^7$Be that gives rise to this hump. Though the dominance of the 
principal value is expected, a hump-like structure in the total 
cross section, due to final state 
$\eta$-nucleus interaction is not so natural to expect. For example, 
such an effect was not observed in our previous studies of the 
$ p d \to p d \eta$ and $p d \to \,^3$He $\eta$ reactions. 
\begin{figure}[h]
\includegraphics[width=7cm,height=7cm]{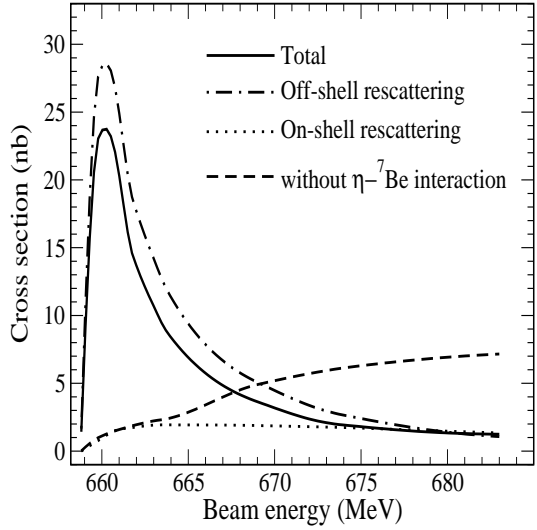}
\caption{\label{fig5:eps1}
Contributions of the plane wave (dashed line), on-shell rescattering 
(dotted line) and off-shell rescattering (dot-dashed line) terms to the 
total cross section (solid line) for the 
${\rm p}\, + \,^6{\rm Li} \, \rightarrow\,  \eta \,+\,^7{\rm Be}$ reaction
with $^7$Be in its ground state ($J = 3/2^-$). The cluster model nuclear wave 
functions are generated using the Woods-Saxon potential.}
\end{figure}

In order to understand this phenomenon, we re-write the principal value 
integral in the second term of Eq. (\ref{prodtmat1})  
(corresponding to off-shell rescattering) in a split form. Let us express the 
integral $\int d\vec{q} = \int d\Omega_q\, \int_0^{\infty} \,q^2\, dq$ 
as $\int d\Omega_q\, \int \,q\, \mu\, dE(q)$ where, 
$q^2 = 2 \mu E(q)$. 
Dropping the spin projection dependence in the 
notation for convenience, we define the rescattering term alone 
(second term in (\ref{prodtmat1})) as a function, 
$G(\vec{k}_p,\vec{k}_{\eta})$ plus the pole term, such that, 
\begin{eqnarray}\label{pvinteg}
\int&&{d\vec{q}\over(2\,\pi)^3}\,{\langle\,\vec{k_{\eta}}\,|\,T_{\eta\,^7Be}
\,|\,\vec{q}\,\rangle \over
E(k_{\eta})\,-\,E(q)\,+\,i\,\epsilon}\,\langle\,\vec{q}\,|\,
T_{p\,^6Li\,\rightarrow\,\eta\,^7Be}\,|\,\vec{k_p}\,\rangle \\
\nonumber 
&&=\,{\cal P}\, \int d\Omega_q\, 
F(\vec{k_p},\vec{k}_{\eta},\hat{q})
\int \,dE(q)\,{f(E(q)) \over E(k_{\eta})\,-\,E(q)} 
\, \,+\, \,{\rm pole}\,\, {\rm term}\, , 
\end{eqnarray}
where the functions, $F$ and $f$ summarize the angle and energy 
dependence (except the energy dependence in the denominator) 
respectively of the vector $\vec{q}$ in the integrand.  
We now break up part of 
the principal value integral into three parts, considering 
the region till and after the pole value as follows: 
\begin{eqnarray}\label{pvalint}
I &=& \int_0^{E(k_{\eta}) - \delta} \,{f(E(q)) \over 
E(k_{\eta})\,-\,E(q)}\, dE(q) \,+\,
\int_{E(k_{\eta}) + \delta}^{2 E(k_{\eta})} \,{f(E(q)) \over 
E(k_{\eta})\,-\,E(q)}\, dE(q) \\ \nonumber
&&+\,
\int_{2 E(k_{\eta})}^{\infty} \,{f(E(q)) \over E(k_{\eta})\,-\,E(q)}\, dE(q)\, 
\,=\,\, I_1\, +\, I_2\, +\, I_3\, .
\end{eqnarray}  
In order to get some idea of the behaviour of the complicated integrals 
in (\ref{pvinteg}) and (\ref{pvalint}) which we perform numerically, we 
try to analyse these integrals analytically in a very simplistic way. 
First, we perform an expansion of $f(E(q))$, such that,
$$f(E(q)) \,\simeq \, f(E(k_{\eta}))\,+\, (E(k_{\eta})\,-\,E(q))\,
f^{\prime}(E(k_{\eta})) \, + \, ... $$ and hence, 
$${f(E(q)) \over E(k_{\eta})\,-\,E(q)} \,\simeq \, 
{f(E(k_{\eta})) \over E(k_{\eta})\,-\,E(q)}\, + \, 
f^{\prime}(E(k_{\eta})) \,+\, ...\,.$$
We further assume that only the first term 
on the right in the above expansion is important. Not always can one 
expect the energy dependence of $f(E(q))$ to be such that we can
write it with only a constant $f(E(k_{\eta}))$ appearing in the first term 
as above. 
If the pole is however close to zero, then the energy region 
$ 0 \to 2\, E(k_{\eta})$ is a small region around the pole 
where $f(E(q))$ may vary little and the expansion
made above becomes a reasonable approximation. 
Since $f(E(k_{\eta}))$ is a constant, when we integrate, we get,
\begin{equation} 
I \,=\,{\rm ln} 
\,(E(k_{\eta})\,-\,E(q)) \,\biggl |_0^{E(k_{\eta})-\delta} \, +\, 
{\rm ln} 
\,(E(k_{\eta})\,-\,E(q)) \,\biggl |_{E(k_{\eta}) +\delta}^{2 E(k_{\eta})} \, 
\,+\, \, I_3\, ,
\end{equation}
where obviously the first two terms cancel. The integral $I$ in 
(\ref{pvalint}) will then be dominated only by $I_3$ whose behaviour 
will decide the shape of the off-shell rescattering term. 
$I_3$ depends on the form of the function $f(E(q))$, 
which is a product of the 
${\rm p}\, + \,^6{\rm Li} \, \rightarrow\,  \eta \,+\,^7{\rm Be}$ 
production t-matrix and the $\eta$-$^7$Be elastic t-matrix. 
The latter is peaked at small $k_{\eta}$ due to the fact that 
the $^7$Be wave function contributes at large $r$. The $I_3$ part of the 
off-shell rescattering term 
which grows as one approaches energies close to threshold gives rise 
to the hump due to FSI in the total cross section. 
The above analysis will not be valid when one is far away from threshold. 
\subsubsection{Single scattering versus multiple scattering to all orders} 
In Fig. 6, for the sake of completeness is shown the cross section 
\begin{figure}[h]
\includegraphics[width=7cm,height=10cm]{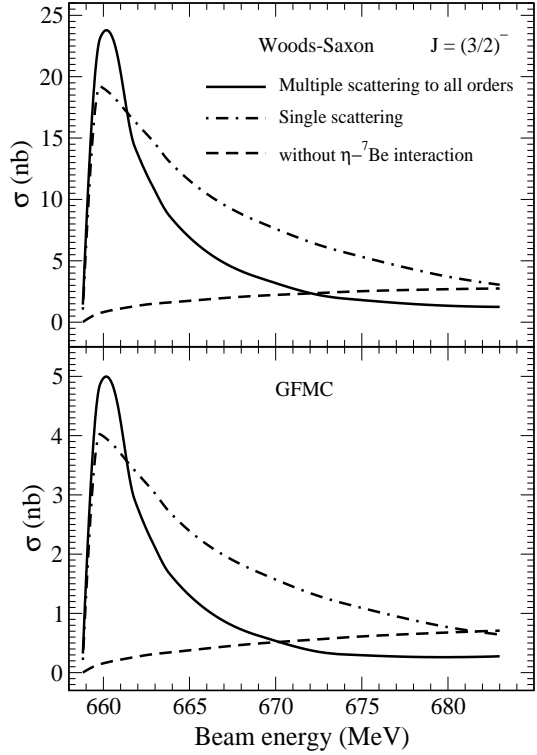}
\caption{\label{fig9:eps1}
Total cross sections evaluated using only the single scattering term 
(dot-dashed) in the $\eta$-$^7$Be interaction, the full calculation 
(solid line) involving $\eta$ meson rescattering and 
propagation inside $^7$Be to all orders and the  
one without 
the inclusion of the $\eta$-$^7$Be interaction (dashed line).  
The calculation is done for the ground state of $^7$Be ($J = 3/2^-$). 
}
\end{figure}
evaluated using only the single scattering terms in the $\eta$-$^7$Be 
t-matrix (i.e. the first terms in Eqs (\ref{eta3He4He2} - \ref{eta3He4He3})) 
as compared to that using the full coupled channel 
t-matrix which includes 
the multiple scattering to all orders (Eqs (\ref{eta3He4He2}-
\ref{eta3He})). 
The dot-dashed line is thus the calculation
corresponding to an $\eta$-$^7$Be interaction where the $\eta$ meson 
scatters once on each of the $^3$He and $^4$He nuclei inside $^7$Be and
then proceeds to become on-shell in the final state of the 
${\rm p}\, + \,^6{\rm Li} \, \rightarrow\,  \eta \,+\,^7{\rm Be}$ reaction. 
The solid line is due to the $\eta$-$^7$Be interaction which involves 
multiple scattering of the $\eta$ on the 
two clusters, $^3$He and $^4$He, and its propagation inside $^7$Be 
in between these scatterings. 

\subsubsection{Angular distributions}
In Fig. 7, we plot the angular distributions for this reaction
\begin{figure}[h]
\includegraphics[width=6cm,height=9cm]{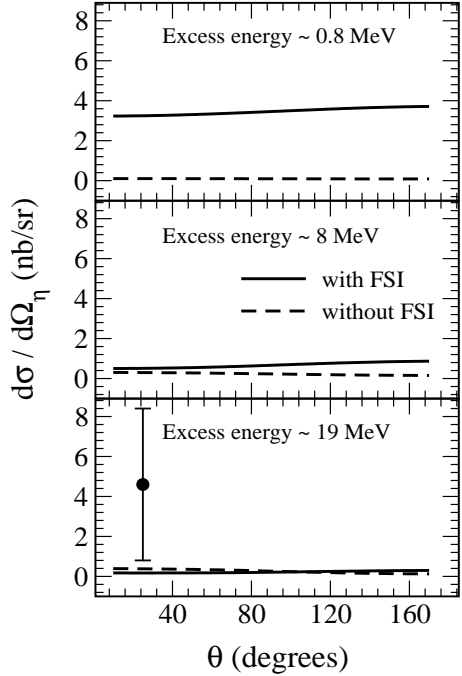}
\caption{\label{fig8:eps1}
Angular distributions for the 
${\rm p}\, + \,^6{\rm Li} \, \rightarrow\,  \eta \,+\,^7{\rm Be}$ reaction
at different excess energies (corresponding to different beam energies). 
The data point is by the Turin group \cite{turin}. Solid and dashed 
lines correspond to the calculations with and without the inclusion 
of the $\eta$-$^7$Be interaction respectively. The calculation is 
done for the $^7$Be ground state and using Woods-Saxon wave functions. 
}
\end{figure}
for the ground state ($J = 3/2^-$) of $^7$Be.  
The angular distributions are nearly isotropic as has always been found in 
the $\eta$ producing reactions near threshold. The FSI between the 
$\eta$ meson and $^7$Be is responsible for raising the magnitudes of the 
cross sections only close to threshold. 
At an excess energy of 19 MeV, the effect of the FSI is highly reduced and the 
theoretical prediction in Fig. 7 is around 0.2 nb/sr as compared to
the Turin data point of 4.6 $\pm$ 3.8 nb/sr. Even if we add up the 
cross sections corresponding to other values of $J$, i.e., 
$J = 1/2^-, 5/2^-$ and $7/2^-$, the theoretical prediction with FSI is 
0.51 nb/sr. Though not very clear at the moment, one could speculate that
the deficit in the theoretical prediction as compared to data is probably
due to some missing reaction mechanism not included in the present 
cluster model approach.

\begin{figure}[h]
\includegraphics[width=9cm,height=12cm]{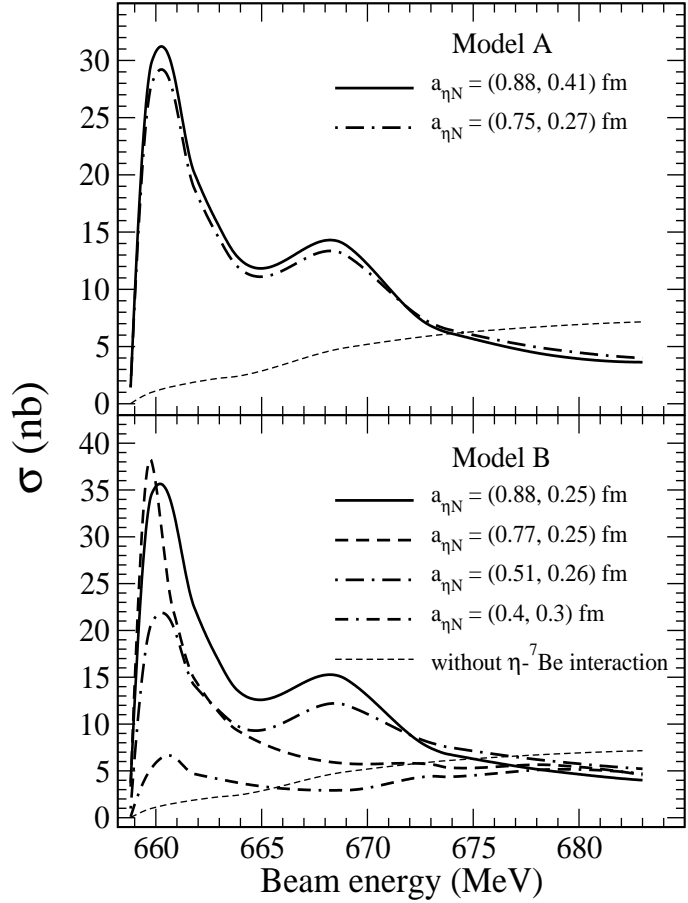}
\caption{\label{fig7:eps1}
Total cross sections for the inclusive reaction 
${\rm p}\, + \,^6{\rm Li} \, \rightarrow\,  \eta \,+\,^7{\rm Be}$. Inclusive 
implies that the cross section for four $J$ values of $^7$Be, namely, 
$1/2^-$, $3/2^-$, $5/2^-$ and $7/2^-$ have been summed. 
The upper and lower panel display results within two different models 
of the $\eta N$ interaction used for evaluating the $\eta$-$^7$Be FSI as 
described in the text. The 
dotted line without any humps corresponds to the calculation without the 
$\eta$-$^7$Be final state interaction.}
\end{figure}
\subsection{Inclusive cross section} 
Having analysed and deciphered the origin of the hump in the 
total cross sections near threshold, we now plot in Fig. 8, the 
sum of the total cross sections,  
$\sum_J\,\,\sigma^J_{{\rm p}\, + \,^6{\rm Li} \, \rightarrow\,  
\eta \,+\,^7{\rm Be}}$, 
where the superscript $J$ refers to the state in which the $^7$Be 
in the final state is produced.
The four different levels of $^7$Be considered here have a different 
value of $J$ and also a different mass. In this sense, the summed cross 
section represents an inclusive cross section of the 
${\rm p}\, + \,^6{\rm Li} \, \rightarrow\,  \eta \,+\,X$ reaction 
with $X$ being any of the low-lying states of $^7$Be. 
We have already seen that within Model A, 
the close to threshold $\eta$-$^7$Be interaction
causes a sharp rise (the hump) in the cross section. Now, in the case
of the $L = 3$ states, as can be seen from Table I, the threshold 
is shifted by about 10 MeV as compared to the $L = 1$ states. As a result,
the $L = 3$ humps appear and contribute at a higher beam energy. 
Since the hump is a close to threshold phenomenon, the $L = 1$ cross sections 
already fall down a lot before the $L = 3$ humps begin. The result is a 
double hump structure in the inclusive 
${\rm p}\, + \,^6{\rm Li} \, \rightarrow\,  \eta \,+\,^7{\rm Be}$ reaction. 

The cross sections evaluated in Model B, have a similar form as 
those in Model A except for two values of the scattering lengths,
namely, (0.77,0.25) fm and (0.4,0.3) fm. For these two values, 
the cross sections do not display a very prominent second hump. 
After a careful reading of Ref. \cite{wycgre} (which lists the 
parameter sets for Model B, lower panel in Fig. 8) one notices that 
the parameter sets for these two particular curves have been obtained 
after removing the $\gamma N \to \eta N$ data from the analysis. 
Thus these two curves are different from all the rest in Fig. 8, 
in the sense that these parameter sets are not constrained by the 
photoproduction data. Model A, however, was used in \cite{fix} to 
calculate the cross sections for the 
$\gamma d \to \eta d$ and $\gamma d \to \eta X$ 
reactions. Hence, Model A and the two curves 
with double humps in Model B are 
constrained by the eta photoproduction data which is probably important 
for the behaviour of the $\eta N$ $t$-matrix 
at energies away from threshold.

\subsection{Comment on possible eta-mesic $^7$Be states} 
Finally, before ending our discussion of the results, in Table II, we 
give the $\eta$-$^7$Be scattering lengths corresponding to different 
values of the $\eta N$ scattering lengths in models A and B. 
The $\eta$-$^7$Be scattering length is evaluated from the elastic 
$\eta$-$^7$Be t-matrix at zero energy as follows:
\begin{equation}
a_{\eta-7Be}\ =\, -\, {\mu \, T_{\eta-7Be}(0,0,0) 
\over 2 \, \pi} \, ,
\end{equation}
where $\mu$ is the reduced mass of the $\eta$-$^7$Be system. 
\begin{table}[h]
\caption{$\eta$-$^7$Be scattering lengths, $a_{\eta -7Be}$, for the 
ground state of $^7$Be,  
corresponding to different values of $a_{\eta N}$.}
\begin{center}
\begin{tabular}{|l|l|l|}
\hline
 &\,\, $a_{\eta N}$ (fm) &\,\, $a_{\eta -7Be}$ (fm)\\
\cline{2-3}
Model A  & $0.75\, + \, i 0.27$& $- \,10.09\, +\, i 8.19$ \\
 & $0.88\, + \, i 0.41$& $- \,9.18\, +\, i 8.53$ \\
\cline{2-3}
\hline
  & $0.88\, + \, i 0.25$& $-\,20.43 \,+\, i 5.43$ \\
Model B  & $0.77\, + \, i 0.25$& $-\,14.52 \,+\, i 14.77$ \\
  & $0.51 \, +\, i 0.26$& $-\, 2.03 \,+\, i 11.29$ \\
  & $0.4\, + \, i 0.3$& $ \, \,\,\,\,0.29 \, +\, i 6.43$\\
\hline
\end{tabular}
\end{center}
\end{table}
The smallest scattering length $a_{\eta N}$ used here, leads to a positive 
real part of the $\eta$-$^7$Be scattering length, 
$a_{\eta -7Be}$. However, this choice of parameters 
(model F in \cite{wycgre}) is mentioned as an unconventional solution 
obtained after dropping the photoproduction data from the fits. 

The first four entries in Table II corresponding to large $a_{\eta N}$, 
display $a_{\eta -7Be}$ to be large with negative real parts. 
Apart from the 
commonly known condition that the real part of the
scattering length should be negative \cite{joachain},  
in the third reference in \cite{bhaleliu}, the 
authors found that the condition for the existence of a bound 
state is that $|a_I| \,< \, |a_R|$ for an eta-nucleus scattering 
length of ($a_R \,+\,i\,a_I$). With these two conditions it seems that the 
first few entries in Table II for the large  
$\eta N$ scattering lengths support the possibility of eta-mesic states. 
In \cite{sibir04}, while investigating the connection between the 
$\eta$-$^3$He scattering lengths and the corresponding binding energies 
and widths, the authors also use the above conditions but 
mention that in reality none of the above can be taken as a 
sufficient condition. 
The bottomline is then that it 
would indeed be premature to base the conclusions only on the 
signs or magnitudes of the scattering lengths.  
One should rather perform a better analysis for the 
search of $\eta$-$^7$Be states using the present $\eta$-$^7$Be 
model and a time delay analysis as in \cite{weme,otherus} or a 
K-matrix analysis as in \cite{wycech} before drawing 
definite conclusions. 

\section{Summary}
The present work aimed at performing a thorough investigation of the effects 
of the $\eta$-$^7$Be interaction in the 
${\rm p}\, + \,^6{\rm Li} \, \rightarrow\,  \eta \,+\,^7{\rm Be}$ reaction 
near threshold. The work was partly motivated by the recent revival of 
interest in this reaction by the COSY-GEM collaboration \cite{cosy} after the 
first measurement in 1993. 
This work also comes as a sequel to our various earlier studies on $\eta$ 
meson production in light nuclei. A two-step model for the 
${\rm p}\, + \, {\rm d} \, \rightarrow \, ^3{\rm He}\, +\, \eta$ reaction
including the $\eta$-$^3$He interaction was tested earlier with data 
by the present authors \cite{we3and4}. This model is used as an input to
develop a cluster model approach for the 
${\rm p}\, + \,^6{\rm Li} \, \rightarrow\,  \eta \,+\,^7{\rm Be}$ reaction
near threshold. The $\eta$-$^7$Be interaction is included in a multiple 
scattering formalism for an $\eta$ scattering on a $^3$He-$^4$He 
cluster inside $^7$Be. The $\eta$-$^3$He and -$^4$He scatterings are
themselves included using few body equations. 
The calculations are done for four low-lying levels of $^7$Be. To the 
best of our knowledge, this is the most detailed study of the 
$\eta$-$^7$Be interaction in the
${\rm p}\, + \,^6{\rm Li} \, \rightarrow\,  \eta \,+\,^7{\rm Be}$ reaction
performed so far. The interesting two hump structure in the summed 
total cross section (summed over $J$) hints toward a very strong near 
threshold effect of the $\eta$-$^7$Be interaction (especially of the
off-shell rescattering of the $\eta$ on $^7$Be) which is worth verifying 
experimentally in future. 
\\
\\
{\bf ACKNOWLEDGEMENTS}\\
The authors gratefully acknowledge V. Jha for the 
help in computing the cluster wave functions using the Woods-Saxon 
potential and K. P. Khemchandani for
her help with the $p d \to \, \eta\, ^3$He programs and many useful 
discussions. The authors are also thankful to the 
anonymous referee for his/her constructive criticism. 
The work done by two of the authors, NJU and BKJ,  
was supported by the Ramanna fellowship, 
awarded by the Department of Science and Technology, Government 
of India, to BKJ.
\\


\begin{thebibliography}{99}
\bibitem{taps}
M. Pfeiffer {\it et al}., Phys. Rev. Lett. {\bf 92}, 252001 (2004); 
G. A. Sokol {\it et al.}, Part. Nucl. Lett. {\bf 102}, 71 (2000). 
\bibitem{bhaleliu}
R. S. Bhalerao and L. C. Liu, Phys. Rev. Lett. {\bf 54} (1985) 865; 
L.C. Liu and Q. Haider, Phys. Rev. C {\bf 34}, 1845 (1986); 
Q. Haider and L.C. Liu, Phys. Rev. C {\bf 66}, 045208 (2002). 
\bibitem{weme}
N. G. Kelkar, K. P. Khemchandani and B. K. Jain,
J. Phys. G:Nucl. Part. Phys. {\bf 32}, L19 (2006); 
N. G. Kelkar, K. P. Khemchandani and B. K. Jain, J. Phys. {\bf G 32}, 1157 
(2006); N. G. Kelkar, Phys. Rev. Lett. {\bf 99}, 210403 (2007). 
\bibitem{others}
S. Wycech, Anthony M. Green and J.A. Niskanen, Phys. Rev. {\bf C 52}, 
544 (1995); V.A. Tryasuchev, Phys. Atom. Nucl. {\bf 60}, 186 (1997) 
(Yad. Fiz. {\bf 60}, 245 (1997)); 
C.Y. Song, X.H. Zhong and L. Li, P.Z. Ning, Europhys. Lett. {\bf 81}, 
42002 (2008); D. Jido, E.E. Kolomeitsev, H. Nagahiro and S. Hirenzaki, 
Nucl. Phys. {\bf A 811}, 158 (2008); H. Nagahiro, D. Jido and S. Hirenzaki, 
arXiv:0811.4516.  
\bibitem{exp3He4He}
J. Berger {\it et al.}, Phys. Rev. Lett. {\bf 61}, 919 (1988); 
B. Mayer {\it et al}., Phys. Rev. C {\bf 53}, 2068 (1996); A. Wro\'nska 
{\it et al.}, Int. J. Mod. Phys. A {\bf 20}, 640 (2005); H., -H., Adam 
{\it et al.}, Phys. Rev. C {\bf 75}, 014004 (2007). 
\bibitem{we3and4}
K.P. Khemchandani, N.G. Kelkar and B.K. Jain, 
Nucl. Phys. {\bf A 708}, 312 (2002); {\it ibid}, 
Phys. Rev. C {\bf 68}, 064610 (2003); 
N.J. Upadhyay, K.P. Khemchandani, B.K. Jain and N.G. Kelkar, 
Phys. Rev. C {\bf 75}, 054002 (2007); K. P. Khemchandani, N. G. Kelkar and 
B. K. Jain, Phys. Rev. C {\bf 76}, 069801 (2007). 
\bibitem{turin}
Scomparin E. {\it et al.}, J. Phys. G: Nucl. Part. Phys. {\bf 19} 
L 51, (1993).
\bibitem{wilkin}
J. S. Al-Khalili, M. B. Barbaro and C. Wilkin, 
J. Phys. G {\bf 19}, 403 (1993).
\bibitem{cosy}
M. Ulicny {\it et al.}, AIP Conf. Proc. {\bf 603}, 543 (2001); 
H. Machner, Acta Phys. Slov. {\bf 56}, 227 (2006); e-Print: nucl-ex/0511034.
\bibitem{berclust}
T. Kajino, T. Matsuse and A. Arima, Nucl. Phys. A {\bf 413}, 323 (1984). 
\bibitem{lithclust}
J. V. Noble, Phys. Rev. C {\bf 9}, 1209 (1974); A. C. Merchant and 
N. Rowley, Phys. Lett. B {\bf 150}, 35 (1985). 
\bibitem{rakit}
S. A. Rakityansky {\it et al.}, Phys. Rev. C {\bf 53}, 2043 (1996); 
S. A. Rakityansky {\it et al.}, Few Body Syst. Suppl. {\bf 9}, 
227 (1995). 
\bibitem{fix}
A. Fix and H. Arenh\"ovel, Eur. Phys. J A {\bf 9}, 119 (2000). 
\bibitem{wycgre}
A. M. Green and S. Wycech, Phys. Rev. C {\bf 71}, 014001 (2005). 
\bibitem{kujaw}
E. Kujawski and E. Lambert, Annals of Physics {\bf 81}, 591 (1973).
\bibitem{crespo}
R. Crespo, A. M. Moro and I. J. Thompson, Nucl. Phys. A {\bf 771}, 26 (2006). 
\bibitem{bely3}
V. B. Belyaev, S. A. Rakityansky and J. Wrzecionko, 
Nucl. Phys. A {\bf 368}, 394 (1981). 
\bibitem{khal}
B. Buck and A. C. Merchant, J. Phys. G {\bf 14} L211 (1988); V. I. Kukulin,  
V. G. Neudatchin and Yu. F. Smirnov, Nucl. Phys. A {\bf 245} (1975) 429. 
\bibitem{for}
J. L. Forest {\it et al.}, Phys. Rev. {\bf C 54}, 646 (1996).
\bibitem{joachain} 
C. J. Joachain 1975 {\it Quantum Collision Theory} (North-Holland), p.288.
\bibitem{sibir04} 
A. Sibirtsev, J. Haidenbauer, J. A. Niskanen and Ulf-G. Meissner, 
Phys. Rev. {\bf C 70}, 047001 (2004); see also J. A. Niskanen, 
arXiv: nucl-th/0508021. 
\bibitem{otherus} 
N. G. Kelkar, M. Nowakowski, K. P. Khemchandani
and S. R. Jain, Nucl. Phys. {\bf A730}, 121 (2004); 
N. G. Kelkar, M. Nowakowski and K. P. Khemchandani, Nucl. Phys. {\bf A724},
357 (2003); {\it ibid}, Mod. Phys. Lett. A {\bf 19}, 2001 (2004). 
\bibitem{wycech}
S. Wycech and A. M. Green, Int. J. Mod. Phys. {\bf A 20}, 637 (2005). 
\end{thebibliography}
\end{document}